\documentclass[fdp,fleqn]{w-art}
\usepackage{times}
\usepackage{w-thm}
\usepackage[]{graphicx}
\begin{document}
\DOIsuffix{DOI 10.1002/prop.201100020}
\Volume{59}
\Issue{7-8}
\Month{mm}
\Year{2011}
\pagespan{3}{}
\Receiveddate{31 January 2011}
\Reviseddate{xxx}
\Accepteddate{1 February 2011}
\Dateposted{March 7, 2011, {\bf Fortschr. Phys. 59, No. 7 – 8, 637 – 645 (2011)}}
\keywords{supersymmetry, superspace, p-branes, supergravity, BPS states, superembedding.}



\title[On superembedding approach and SO(32) heterotic 5-brane]{On superembedding approach and its possible application in search for
SO(32) heterotic five-brane equations}


\author[I. A. Bandos]{Igor A. Bandos\inst{1,2}%
  \footnote{Corresponding author\quad E-mail:~\textsf{igor\_bandos@ehu.es},
            }}
\address[\inst{1}]{Department  of
Theoretical Physics and History of Science, University of the Basque Country
 EHU/UPV,  P.O.Box 644,  48080 Bilbao, Spain}
\address[\inst{2}]{IKERBASQUE, Basque Foundation for Science, 48011, Bilbao, Spain}
\begin{abstract}

We elaborate the superembedding description of 'simple' $D = 10$, $N = 1$ five-brane and discuss the  possible application of the superembedding approach in search for the equations of motion of the mysterious $SO(32)$ heterotic five brane on this basis.


\end{abstract}
\maketitle                   





\section{Introduction}

Supersymmetric extended objects called {\it super-$p$-branes}, or shorter, {\it $p$-branes}, play an important role in search for quantum theory of gravity and for unified theory of fundamental interactions in the framework presently known under the name of String/M-theory \cite{M-theory}. The set of $p$-branes includes the $D=10$ fundamental superstring in its type II \cite{G+S84} and heterotic \cite{Gross+H+M1984} versions,  as well as M2-brane, also known as 11D supermembrane \cite{BST1987}. Other important objects are  $D=10$ type II Dirichlet p-branes (D$p$-branes) described by the Dirac--Born-Infeld plus Wess--Zumino action \cite{M2->D2,Dpac,B+T=Dpac,bst97} and the M-theory 5-brane or M5-brane whose action was constructed in \cite{blnpst,schw5} with the use of the so--called PST (Pasti--Sorokin--Tonin) technique \cite{pst,pst=bM5}. Interestingly enough, the M5-brane equations of motion were obtained in \cite{hs2} in the frame of superembedding approach \cite{bpstv,hs96,Dima99} two  months before the action was found in \cite{blnpst,schw5}.

The ground states of $p$--branes can be also described by supersymmetric  solutions of the $D=11$ and $D=10$ supergravity equations \cite{Duff94,Stelle98}. The worldvolume actions can be considered as effective action for the excitations over such a solitonic solutions. The worldvolume and supergravity approaches to superbranes are complementary and   an achievement in one of them often stimulates a development of the other. For instance, the supersymmetric solutions of type II supergravity equations corresponding to the ground states of D$p$--branes were found in early nineties \cite{Duff:1993ye} and stimulated the search for the corresponding worldvolume actions; the relation of these solutions and actions with  Dirichlet $p$-branes defined as surfaces where the open string can have its ends \cite{PolchD} was appreciated several years latter \cite{PolchD} and helped to find the worldvolume actions of D$p$--branes \cite{M2->D2,Dpac}. Notice that the superembedding approach to D$p$-branes was proposed in \cite{hs96} where it was shown (using linear approximation) that the superembedding equation results in D$p$-brane equations of motion. This was done before the action for generic D$p$-brane was found in \cite{Dpac} (although after the derivation of a particular representative of this family, the D$2$-brane action, by dimensional reduction of the M2-brane action in \cite{M2->D2}).

Thus the  superembedding approach shown its usefulness as a tool to obtain equations of motion in the cases where the construction of the action principle met some difficulties. This suggested to use it to search for equations of motion of the  systems of multiple D$p$--brane, the program proposed and realized for the case of multiple D$0$-brane in \cite{IB09:D0} and for the case of multiple M$0$-branes in \cite{mM0=PLB,mM0=PRD}. (The cases of multiple $p$-brane systems with $p>1$ are presently under investigation).

The main message of this contribution is the proposition to apply the superembedding approach in search for the equations of motion of the heterotic 5-branes \cite{Witten:1995gx} (see \cite{Lechner:2010ti} and refs. therein). We present here the candidate basic equations for the description of the SO(32) heterotic 5-brane  in the frame of  the superembedding approach. The selfconsistency of these equations is under study now so that the present contribution can be considered as a progress report.

We begin here by elaborating the superembedding approach to `simple' $D=10$, ${\cal N}=1$ five-brane, the object  which was described already in the first 'brane scan' of \cite{AETW}. This provides us  with the background necessary to formulate our proposition for the basic equations of the superembedding approach to the SO(32) heterotic five brane, and also by the complete description of the geometry of the worldvolume superspace where the proposed equations are formulated.

%

\section{Superembedding approach to `simple' $D=10\,$, $\;{\cal N}=1$ five-brane}



Superembedding approach \cite{bpstv,hs96,hs2,bst97,Dima99} provides the superfield description of the worldvolume dynamics of supersymmetric extended objects, $p$-branes. The $p+1$ dimensional worldvolume $W^{p+1}$ is extended in it till the worldvolume superspace ${\cal W}^{(p+1|{n\over 2})}$ with $p+1$ bosonic and  ${n\over 2}$ fermionic dimensions, where $n$ is the number of fermionic directions of the target superspace $\Sigma^{(D|n)}$ (in particular, $n=32$ for $D=11$ and type II $D=10$ theories, while for $D=10$, ${\cal N}=1$ theories $n=16$)\footnote{See  \cite{Dima99} for the review of versions of superembedding approach and STV approach in superspaces with less then $n/2$ fermionic coordinates.}.

Hence, in  the case of simple $D=10$, ${\cal N}=1$ five-brane, as well as in the case of heterotic five-branes, the suitable worldvolume superspace must have 6 bosonic and 8 fermionic dimensions.
We denote the local bosonic and fermionic coordinates of such a worldvolume  superspace
${\cal W }^{(6|8)}$ by  $\xi^m$ and $\eta^{\mu}$, respectively, and use the notation  $\zeta^{\cal M}$ for their complete set,
\begin{eqnarray}
\label{W:zeta=} \zeta^{\cal M}=(\xi^m\; , \; \eta^{\mu}) \; , \qquad \eta^{\mu}\eta^{\nu}=- \eta^{\nu}\eta^{\mu}\; , \qquad m=0,1,...,5\; , \qquad \mu=1,...,8\; ,
\end{eqnarray}

The embedding of the 5-brane worldvolume superspace ${\cal W}^{(6|8)}$ into the  $D=10$, ${\cal N}=1$ superspace $\Sigma^{(10|16)}$ can be described in terms of
coordinate functions $\hat{Z}^{\underline{\cal M}}(\zeta)= (\hat{x}{}^{
\underline{m}}(\zeta)\, , \hat{\theta}^{ \underline{\mu}}(\zeta))$,
\begin{eqnarray}
\label{WinS} & {\cal W }^{(6|8)}\in \Sigma^{(10|16)} : \quad
Z^{\underline{\cal M}}= \hat{Z}^{\underline{\cal M}}(\zeta) \qquad \Leftrightarrow \qquad \begin{cases}{x}{}^{
\underline{m}}= \hat{x}{}^{
\underline{m}}(\zeta)\;  , \quad \underline{m}=0,1,...,9\; , \cr  {\theta}^{ \underline{\mu}}= \hat{\theta}^{ \underline{\mu}}(\zeta)\; , \qquad \underline{\mu}=1,...,16\; . \end{cases} \;  . \qquad
\end{eqnarray}
In the case of 'simple' five--brane, the dependence of the coordinate functions on the local coordinates of ${\cal W }^{(6|8)}$, Eq. (\ref{W:zeta=}), is determined by the superembedding equation.

\subsection{{\it Superembedding equation for a $D=10$ ${\cal N}=1$ five-brane}}

To write this equation we must introduce first the supervielbein forms of the target superspace $\Sigma^{(10|16)}$
\begin{eqnarray}
\label{EA=(Eb,Ef)}
{E}^{ \underline{\cal A}}:=
dZ^{\underline{\cal M}}E_{\underline{\cal M}}{}^{\underline{\cal A}}(Z)=
 ({E}^{ \underline{a}}, {E}^{ \underline{\alpha}})\; ,  \qquad
 \underline{a}=0,1,...,9\; , \qquad \underline{\alpha}=1,2,...,16 \; .  \qquad
\end{eqnarray}
Their set includes the bosonic vector form ${E}^{ \underline{a}}$ and the fermionic spinor 1-form ${E}^{ \underline{\alpha}}$. The pull--back $\hat{E}^{\underline{\cal A}}:=
d\hat{Z}{}^{\underline{\cal M}}E_{\underline{\cal M}}{}^{\underline{\cal A}}(\hat{Z})$ of these supervielbein forms to ${\cal W }^{(6|8)}$ can be decomposed on the basis of supervielbein forms of ${\cal W }^{(6|8)}$, the set of which,
\begin{eqnarray}
\label{W:eA=}
   e^{\cal A}:= (e^a\; , \; e^{\alpha {A}}):= d\zeta^{{\cal M}} e_{{\cal M}}{}^{{\cal A}}(\zeta) \qquad
   a=0,1,...,5\; , \quad \alpha =1,2,3,4\; , \quad A=1,2\; .
 \end{eqnarray}
includes the bosnonic 6-vector one-form $e^a=d\zeta^{{\cal M}} e_{{\cal M}}{}^a(\zeta)$ and the $SU(2)$ doublet of $SO(1,5)$- spinor fermionic forms $ e^{\alpha {A}}=d\zeta^{{\cal M}} e_{{\cal M}}{}^{\alpha {A}}(\zeta)$.

The generic decomposition reads
\begin{eqnarray}
\label{hEA=ehE}
\hat{E}^{ \underline{\cal A}}:=
d\hat{Z}{}^{\underline{\cal M}}E_{\underline{\cal M}}{}^{\underline{\cal A}}(\hat{Z})=
 e^{\cal B} {\cal D}_{\cal B}\hat{Z}{}^{\underline{\cal M}}E_{\underline{\cal M}}{}^{\underline{\cal A}}(\hat{Z}) =:
 e^{\cal B}\hat{E}_{\cal B}{}^{ \underline{\cal A}} = e^{\beta B}\hat{E}_{\beta B}{}^{ \underline{\cal A}}+
 e^{b}\hat{E}_{b}{}^{ \underline{\cal A}}\; .  \qquad
\end{eqnarray}
The superembedding equation states that the pull--back of the bosonic supervielbein of the target superspace to ${\cal W }^{(6|8)}$ has no fermionic projection
\footnote{In Eqs. (\ref{hEA=ehE}) and (\ref{SembEq=}) ${\cal D}_{\cal A}= ({\cal D}_{a}, {\cal D}_{\alpha A})$ denotes (bosonic and fermionic) covariant derivatives for the scalar worldvolume superfields on the (generically curved) worldvolume superspace ${\cal W}^{(6|8)}$, $\; e_{\cal A}^{\,\cal M}(\zeta) \partial_{\cal M}$. Below we will use the same symbol ${\cal D}_{\cal A}$ to denote the $SO(1,5)\otimes SO(4)$ covariant derivatives, which may include corresponding connections. This symbol is also used to denote $SO(1,5)\otimes SO(4)\otimes SU(2)$ covariant derivative, when speaking about heterotic five-brane in the concluding section.},
\begin{eqnarray}
\label{SembEq=}
  \hat{E}_{\beta B}{}^{\underline{a}}:=  {\cal D}_{\beta B}\hat{Z}^{\underline{\cal M}}\;  E_{\underline{\cal M}}{}^{\underline{a}}(\hat{Z}) = 0\; .
\qquad
\end{eqnarray}

\subsection{{\it Moving frame, spinor moving frame, equivalent forms of the superembedding equation and conventional constraints}}
Using (\ref{hEA=ehE}) we can equivalently write the superembedding equation in the form $\hat{E}^{ \underline{a}}=
e^b \hat{E}_b^{ \underline{a}}$. If the bosonic body of the worldvolume superspace is not degenerate, then $6$ ten-vectors $\hat{E}_b^{ \underline{a}}$ are independent for any choice of the worldvolume supervielbein. This is still indefinite. To define the  bosonic worldvolume supervielbein forms as induced by the (super)embedding, we state that  the set of vectors $u_b^{ \underline{a}}=\hat{E}_b^{ \underline{a}}$ is orthogonal and normalized,
\begin{eqnarray}
\label{Ea=SembEq}
  \hat{E}^{\underline{a}} =  e^b u_b^{\underline{a}}\; , \qquad u_{a\, \underline{a}}u_b^{\underline{a}}=\eta_{ab}=diag(+,-,-,-,-,-)\; .  \qquad
\end{eqnarray}
Indeed, this implies that \begin{eqnarray}
\label{ea=Eua}e^a= \hat{E}^{\underline{a}} u^a_{\underline{a}}
 \; ,  \qquad
\end{eqnarray}
 {\it i.e.} that the worldvolume vielbein is induced by (super)embedding.

The identification $u_b^{ \underline{a}}=\hat{E}_b^{ \underline{a}}$ implies that 6 vectors $u_b^{\underline{a}}$ are tangential to the worldvolume superspace ${\cal W }^{(6|8)}$. Actually, it is convenient to complete their set till {\it moving frame} by introducing four spatial 6-vectors $u_{B\check{B}}^{\underline{a}}$ orthogonal to them and normalized on (-2),\footnote{
$SO(4)=SU(2)\times SU(2)$, so that $u_{B\check{B}}$ is a vector of $SO(4)$ which is the structure group of normal bundle over ${\cal W}^{(6|8)}$.}
\begin{eqnarray}
\label{I=uu}
  \delta_{\underline{b}}{}^{\underline{a}}= u_{\underline{b}}{}^c{}u_c{}^{\underline{a}}   - {1\over 2} u_{\underline{b}}^{A\check{B}}{}u_{A\check{B}}{}^{\underline{a}}\; , \qquad u^c_{ \underline{a}}u^{B\check{B}\underline{a}}=0 \; , \qquad u^{A\check{A}}_{\underline{a}}u^{B\check{B}\underline{a}}= - 2 \epsilon^{AB} \epsilon^{\check{A}\check{B}} \; .  \qquad
\end{eqnarray} These vectors are orthogonal to ${\cal W }^{(6|8)}$ as far as  $\hat{E}_c{}^{\underline{a}}u^{B\check{B}}_{\underline{a}}=u_c{}^{\underline{a}}u^{B\check{B}}_{\underline{a}}=0$. Taken together, the original form of superembedding equation (\ref{SembEq=}) and the last equality imply
\begin{eqnarray}
\label{EAdA=0}
  \hat{E}^{A\check{A}}:=\hat{E}^{\underline{a}} u^{A\check{A}}_{\underline{a}}=0 \; ,  \qquad
\end{eqnarray}
which, thus, can be considered as an equivalent form of the superembedding equation for $D=10$ 5-brane.

To define the fermionic supervielbein induced by superembedding, it is convenient to introduce the auxiliary {\it spinor moving frame superfields} (see {\it e.g.} \cite{IB09:D0,mM0=PRD,bpstv,bst97} and refs. therein). In the case of $D=10$  super-five-brane these highly constrained superfields  can be defined as two rectangular  blocks of a $Spin(1,9)$ valued matrix ({\it spinor moving frame matrix})
\begin{eqnarray}
\label{VinSpin}V_{\underline{\alpha}}{}^{(\underline{\beta})}= (v_{\underline{\alpha}}{}^{{\beta}B}, v_{\underline{\alpha}}{}_{\beta}^{B})\in Spin(1,9)\; ,  \qquad \beta= 1,...,4\; , \qquad A=1,2\; , \qquad \check{A}=1,2\, .
\end{eqnarray}
The spinor moving frame variables $v_{\underline{\alpha}}{}^{{\beta}B}$, $v_{\underline{\alpha}}{}_{\beta}^{B}$ (also called spinorial Lorentz harmonics,  see \cite{IB09:D0,mM0=PRD,bpstv,bst97} and refs. therein) are
related to the moving frame vectors by the following square--root--type relations
\begin{eqnarray}
\label{vv=us}
v^{\alpha A}\tilde{\sigma}_{\underline{a}}v^{\beta B}= \epsilon^{AB} \tilde{\gamma}_b^{\alpha\beta}u_{\underline{a}}{}^b \; , \qquad v_{\alpha}^{\check A}\tilde{\sigma}_{\underline{a}}v_{\beta}^{\check B}= -\epsilon^{\check{A}\check{B}} {\gamma}_{b\alpha\beta}u_{\underline{a}}{}^b \; , \qquad v^{\alpha A}\tilde{\sigma}_{\underline{a}}v_{\beta}^{\check B}= \delta_{\beta}^{\alpha} \; , \qquad
\\
 u^{b}_{\underline{a}} \sigma^{\underline{a}}_{\underline{\alpha}\underline{\beta}} = \epsilon_{AB}  v_{(\underline{\alpha}}^{ A}\gamma^a v_{\underline{\beta})}^{ B} - \epsilon_{\check{A}\check{B}} v_{(\underline{\alpha}}^{\check{A}} \tilde{\gamma}{}^{b} v_{|\underline{\beta})}^{\check{A}}\; , \qquad u^{A\check{A}}_{\underline{a}} \sigma^{\underline{a}}_{\underline{\alpha}\underline{\beta}} = 4 v_{(\underline{\alpha}}{}^{\gamma A}v_{\underline{\beta}\,)\gamma}{}^{\check A}\; .
 \end{eqnarray}
In this relations  $v_{(\underline{\alpha}}^{ A}\gamma^a v_{\underline{\beta})}^{B}= v_{(\underline{\alpha}|}^{\gamma A}\gamma^a_{\gamma\delta}v_{|\underline{\beta})}^{\delta B}$, $\quad
  v_{(\underline{\alpha}}^{\check{A}} \tilde{\gamma}{}^{b} v_{\underline{\beta})}^{\check{B}}:= v_{(\underline{\alpha}|}{}_\gamma^{\check{A}} \tilde{\gamma}{}^{b\gamma\delta} v_{|\underline{\beta})}{}_\delta^{\check{B}}$,  where $\gamma^a_{\gamma\delta}$ and  $\tilde{\gamma}{}^{b\gamma\delta}$
are $d=6$ Klebsh--Gordan coefficients ($d=6$ counterparts of the relativistic Pauli matrices), which obey $\gamma^a_{\alpha\beta}=- \gamma^a_{\alpha\beta}= {1\over 2}\epsilon_{\alpha\beta\gamma\delta}\tilde{\gamma}^{a\gamma\delta}$ and $\gamma^{(a}\tilde{\gamma}{}^{b)}=\eta^{ab}$, while $\sigma^{\underline{a}}_{\underline{\alpha}\underline{\beta}} = + \sigma^{\underline{a}}_{\underline{\beta}\underline{\alpha}}$ and $\tilde{\sigma}{}^{\underline{a}\underline{\alpha}\underline{\beta}} = + \tilde{\sigma}{}^{\underline{a}\underline{\beta}\underline{\alpha}}$ are $D=10$ relativistic Pauli matrices, which obey $\sigma^{(\underline{a}}\tilde{\sigma}{}^{\underline{b})}=\eta^{(\underline{a}\underline{b})}=diag (+,-,-,-,-,-,-,-,-,-)$.

One can also write similar representation for the moving frame vectors in terms of the elements of the inverse moving frame matrix
 \begin{eqnarray}
\label{V-1inSpin}V_{\; (\underline{\beta})}^{-1\, \underline{\alpha}}= \left(\begin{matrix} v_{{\beta}B}{}^{\underline{\alpha}}\cr v_B^{\beta \underline{\alpha}}\end{matrix}\right) \in Spin(1,9)\; ,  \qquad \beta= 1,...,4\; , \qquad A=1,2\; , \qquad \check{A}=1,2\, .
\end{eqnarray}
 For instance,
 \begin{eqnarray}
 \label{uts=vv}
 u^{b}_{\underline{a}} \tilde{\sigma}^{\underline{a}\underline{\alpha}\underline{\beta}} = \epsilon^{AB}  v^{(\underline{\alpha}}_{ A}\tilde{\gamma}^b v^{\underline{\beta})}_{ B} - \epsilon^{\check{A}\check{B}} v^{(\underline{\alpha}}_{\check{A}} {\gamma}{}^{b} v^{\underline{\beta})}_{\check{B}}\; , \qquad u_{A\check{A}}{}^{\underline{a}}\tilde{\sigma}_{\underline{a}}{}^{\underline{\alpha}\underline{\beta}}  = -4  v_{\check A}{}^{\gamma(\underline{\alpha}}v_{\gamma A}{}^{\underline{\beta})}\; .
\end{eqnarray}
The following relations between elements of spinor moving frame matrix and of its inverse are useful
\begin{eqnarray}
 \label{vs=ugv}
v_{\beta B}{}^{\underline{\beta}}{\sigma}^{\underline{a}}_{\underline{\beta}\underline{\alpha}} = u_{b}^{\; \underline{a}} {\gamma}{}^{b}_{\beta \gamma}\epsilon_{BC} v_{\underline{\alpha}}^{\gamma C} - v_{\underline{\alpha}}{}_\beta^{\check B} u_{B \check B}{}^{\underline{a}}
\, , \quad {\sigma}^{\underline{a}}_{\underline{\beta}\underline{\alpha}}v_{\check A}{}^{\gamma \underline{\alpha}} = - \epsilon_{\check{A}\check{B}}  v_{\underline{\beta}}{}_\beta^{\check B}  \tilde{\gamma}{}^{b \beta \gamma} u_{b}^{\; \underline{a}} - v_{\underline{\beta}}^{\gamma B} u_{B \check A}{}^{\underline{a}}
\, . \quad
\end{eqnarray}
In particular, they help in deriving the following useful identities
\begin{eqnarray}
 \label{vs3v'=}
 v_{\beta B}\sigma^{\underline{a}_1\underline{a}_2\underline{a}_3}v_{\check A}^\gamma= - 3 u_b^{[\underline{a}_1} u_c^{\underline{a}_2}u_{B \check A}^{\underline{a}_3]} \gamma^{bc}{}_\beta{}^\gamma - u_{C \check A}^{[\underline{a}_1} u^{\underline{a}_2| C \check C}u_{B \check C}^{|\underline{a}_3]} \delta_\beta{}^\gamma
 \, , \quad \\
 \label{vs3v=}
 v_{\beta B}\sigma^{\underline{a}_1\underline{a}_2\underline{a}_3}v_{\gamma C}= - u_a^{[\underline{a}_1} u_b^{\underline{a}_2}u_{c}^{\underline{a}_3]} \gamma^{abc}{}_{\beta\gamma} - 3 u_{a}^{[\underline{a}_1} u_B^{\underline{a}_2|\check C}u_{C \check C}^{|\underline{a}_3]} \gamma^{a}{}_{\beta\gamma}
 \, . \quad
\end{eqnarray}

The worldvolume fermionic supervielbein induced by superembedding is then defined by
\begin{eqnarray}
 \label{ef=Efv}
 e^{\alpha A} = \hat{E}^{\underline{\alpha}} v_{\underline{\alpha}}{}^{\alpha A}\; . \qquad
\end{eqnarray}
It is convenient to define the $SO(1,5)$ and $SO(4)$ connections  by specifying the $SO(1,5)\otimes SO(4)$ covariant derivative acting on the moving frame vectors to be of the form (see \cite{bpstv})
\begin{eqnarray}
\label{Du=uOm}   {\cal D}u_{\underline{b}}^{\; a} = {1\over 2} u_{\underline{b}A\check A}\Omega^{a\, A\check A} \; , \qquad  {\cal D}u_{\underline{b}}^{A\check A}= {1\over 2} u_{\underline{b}a}\Omega^{a\, A\check A} \; . \qquad
\end{eqnarray}
The covariant one-forms  $\Omega^{a\, A\check A} $ can be considered as generalizations of the $SO(1,9)\over {SO(1,5)\otimes SO(4)}$ Cartan forms for the case of local $SO(1,9)$ symmetry of curved target superspace of the five-brane model. The derivatives of spinor moving frame variables read
\begin{eqnarray}
\label{Dv-q=}  {\cal D}v_{\underline{\alpha}}^{\beta B}={1\over 2} v_{\underline{\alpha}}{}_\gamma^{\check{A}} \tilde{\gamma}_a^{\gamma\beta} \epsilon_{\check{A}\check{B}} \Omega^{a\, B\check B}\; , \qquad {\cal D}v_{\underline{\alpha}}{}_\beta^{\check{B}}={1\over 2} v_{\underline{\alpha}}^{\gamma A} \gamma^a_{\gamma\beta} \epsilon_{{A}{B}} \Omega^{a\, B\check B}\; . \qquad
\end{eqnarray}

The worldvolume curvature two form, $r^{ab}=-r^{ba}$ and the curvature of normal ($SO(4)=SU(2)\otimes SU(2)$) bundle over the worldvolume superspace, ${\cal F}_{B} {}^{A}$ and ${\cal F}_{\check B} {}^{\check A}$,   can be now defined by Ricci identities
\begin{eqnarray}
\label{DDu=uOm}   {\cal D} {\cal D}u_{\underline{b}}^{\; a} = \hat{R}_{\underline{b}}{}^{\underline{a}} u_{\underline{a}}^{\; a} -  u_{\underline{a}}^{\; b} r_{b}^{\; a}\; , \qquad
  {\cal D}{\cal D}u_{\underline{b}}^{A\check A}= \hat{R}_{\underline{b}}{}^{\underline{a}} u_{\underline{a}}^{A\check A} -  u_{\underline{a}}^{B\check A} {\cal F}_B{}^A-  u_{\underline{a}}^{A\check B}  {\cal F}_{\check B} {}^{\check A}  \; , \qquad
\end{eqnarray}
where $\hat{R}_{\underline{b}}{}^{\underline{a}} $ is the pull--back of the SO(1,9) curvature of target superspace (see Eq. (\ref{Rab=10D}) below).

Substituting (\ref{Du=uOm}) for $ {\cal D}u_{\underline{b}}^{\; a}$ and ${\cal D}u_{\underline{b}}^{A\check A}$ in these equations we find the following superfield generalization of the Peterson--Codazzi, Gauss and Ricci equations (see \cite{bpstv})
\begin{eqnarray}
\label{PC+G=Eqs}  & D\Omega^{a\, A\check A}= \hat{R}{}^{a\, A\check A}  \; , \qquad  r^{ab}=  \hat{R}^{ab} + {1\over 2} \Omega^{a}_{A\check A}\wedge \Omega^{b\, A\check A}\; , \qquad  \\ \label{R-Eq}
& {\cal F}_B{}^A= {1\over 4}  \hat{R}{}_{B\check B}^{\; A\check B}+ {1\over 4} \Omega_{B\check B}\wedge \Omega^{b\, A\check B} \; , \qquad {\cal F}_{\check B} {}^{\check A}= {1\over 4}\hat{R}{}_{B\check B}^{\; B\check A}+ {1\over 4} \Omega_{b B\check B}\wedge \Omega^{b\, B\check A}
 \; ,
\end{eqnarray}
where $\hat{R}{}^{a\, A\check A} := \hat{R}^{\underline{a}\underline{b}} u_{\underline{a}}^{a}u_{\underline{b}}^{A\check A}$, $\hat{R}{}^{a\, b} := \hat{R}^{\underline{a}\underline{b}} u_{\underline{a}}^{a}u_{\underline{b}}^{b}$ and $\hat{R}{}_{B\check B}^{\;  A\check A} := \hat{R}^{\underline{a}\underline{b}} u_{\underline{a} B\check B}u_{\underline{b}}^{A\check A}$.

\subsection{{\it From the superembedding equation to equations of motion for simple super-five-brane}}

The selfconsistency conditions for the superembedding equation can be collected in the differential form equation \begin{eqnarray}
\label{DSemb=} 0={\cal D}\hat{E}^{A\check{A}}= \hat{T}^{\underline{a}} u^{A\dot{A}}_{\underline{a}} + \hat{E}^{\underline{a}} \wedge {\cal D} u^{A\check{A}}_{\underline{a}}
\; ,
\end{eqnarray} where $\wedge $ denotes the wedge product of differential forms,
\begin{eqnarray}
\label{wedge}
e^a \wedge e^b= - e^a \wedge e^b\; , \qquad e^a\wedge  e^{\alpha A}= - e^{\alpha A}\wedge e^a\; , \qquad e^{\alpha A}\wedge e^{\beta B}= + e^{\beta B}\wedge e^{\alpha A} \; , \qquad
\end{eqnarray}
$ {\cal D}:=  e^{\cal A} {\cal D}_{\cal A}= e^a{\cal D}_a+   e^{\alpha A}{\cal D}_{\alpha A}$  denotes the $SO(1,5)\otimes SO(4)$ covariant derivative and $\hat{T}^{\underline{a}}$ is the pull--back to  ${\cal W}^{(6|8)}$ of the bosonic torsion two-form of the target superspace ${\Sigma}^{(10|16)}$. This is defined as $SO(1,9)$ covariant exterior derivative of the bosonic supervielbein form, $T^{\underline{a}}:= DE^{\underline{a}} := dE^{\underline{a}}- E^{\underline{b}}\wedge \omega_{\underline{b}}{}^{\underline{a}}$. The   $D=10$, ${\cal N}=1$ supergravity constraints implies that \cite{10D1N=SG}
\begin{eqnarray}
\label{Ta=10D} & T^{\underline{a}}:= DE^{\underline{a}} =
-i{E}^{\underline{\alpha}} \wedge  {E}^{\underline{\beta}} \sigma^{\underline{a}}_{\underline{\alpha}\underline{\beta}}\; , \qquad
\end{eqnarray}
where $\sigma^{\underline{a}}_{\underline{\alpha}\underline{\beta}}= \sigma^{\underline{a}}_{\underline{\beta}\underline{\alpha}}$ are $D=10$ Pauli matrices. Studying Bianchi identities with (\ref{Ta=10D}), one finds \cite{10D1N=SG}\footnote{See \cite{10D1N=SG+} for modifications of the constraints to account for anomalies/modifications of the Bianchi identities for the 3-form and 7-form field strengths. We will not need in such a more complicated constraints for our discussion here.}
\begin{eqnarray}
\label{Tf=10D} & T^{\underline{\alpha}}:= DE^{\underline{\alpha}} =
{i\over 4} {E}^{\underline{b}} \wedge  {E}^{\underline{\beta}} (\sigma^{\underline{a}_1\underline{a}_2\underline{a}_3}\sigma_{\underline{b}})_{\underline{\beta}}{}^{\underline{\alpha}}
h_{\underline{a}_1\underline{a}_2\underline{a}_3} + {1\over 2}{E}^{\underline{b}} \wedge  {E}^{\underline{a}} T_{\underline{a}\underline{b}}{}^{\underline{\alpha}}
\; , \qquad
 \\
\label{Rab=10D} & R^{\underline{a}\underline{b}}:=d\omega^{\underline{a}\underline{b}}- \omega^{[\underline{a}|\underline{c}}\wedge \omega_{\underline{c}}{}^{|\underline{b}]}  = {1\over 2}
{E}^{\underline{\alpha}} \wedge  {E}^{\underline{\beta}} \left( \sigma^{\underline{a}_1\underline{a}_2\underline{a}_3\underline{a}\underline{b}} h_{\underline{a}_1\underline{a}_2\underline{a}_3} - 6 h^{\underline{a}\underline{b}\underline{c}}\sigma_{\underline{c}}\right)_{\underline{\alpha}\underline{\beta}}
 + \nonumber \\ & + {E}^{\underline{c}} \wedge  {E}^{\underline{\beta}}
 \big[
-iT^{\underline{a}\underline{b}\underline{\beta}}\sigma_{\underline{c}}{}_{\underline{\beta}\underline{\alpha}} + 2i T_{\underline{c}}{}^{[\underline{a} \, \underline{\beta}}
\sigma^{\underline{b}]}{}_{\underline{\beta}\underline{\alpha}} \, \big]  + {1 \over 2} E^{\underline{d}} \wedge E^{\underline{c}}
R_{\underline{c}\underline{d}}{}^{\underline{a}\underline{b}} \;
\end{eqnarray}
The leading component of the tensorial superfield $h_{\underline{a}_1\underline{a}_2\underline{a}_3}= h_{[\underline{a}_1\underline{a}_2\underline{a}_3]}$, involved in the solution (\ref{Tf=10D}) and (\ref{Rab=10D}) of the supergravity constraints, is related to the field strength of the two-form gauge field (Kalb-Ramond gauge field $B_{ab}=B_{[ab]}$) which enters the $D=10$, ${\cal N}=1$ supergravity multiplet.

Using (\ref{Ta=10D}) one can present the first term in the {\it r.h.s.} of (\ref{DSemb=}) as
$ -i{E}^{\underline{\alpha}} \wedge  {E}^{\underline{\beta}}
\sigma^{\underline{a}}_{\underline{\alpha}\underline{\beta}} u^{A\dot{A}}_{\underline{a}}$.
Then, using the properties (\ref{vv=us}) and (\ref{Du=uOm}) of the moving frame variables
and conventional constraints $e^a= \hat{E}^{\underline{a}} u^a_{\underline{a}}$
(see Eq. (\ref{Ea=SembEq})) and (\ref{ef=Efv}), we present  Eq. (\ref{DSemb=}) in the form of
$-4i  e^{\alpha A} \wedge \hat{E}^{\check{A}}_\alpha +
e_b \wedge
\Omega^{b\, A\dot{A}} =0 $.
This equation implies that the fermionic form $\hat{E}^{\check{A}}_\alpha¨:=\hat{E}^{\underline{\alpha}}v_{\underline{\alpha}}{}^{\check{A}}_\alpha$ does not have fermionic projection, so that $\hat{E}^{\check{A}}_\alpha = e^a \chi_a{}^{\check{A}}_\alpha$, and also expresses the generalized Cartan form $\Omega^{b\, A\dot{A}} $ in terms of the fermionic superfield $\chi_a{}^{\check{A}}_\alpha:= \hat{E}_a{}^{\check{A}}_\alpha $ and four symmetric tensor superfields $K^{ab\; A\check{A}}=K^{ba\;  A\check{A}}$,
\begin{eqnarray}\label{DSemb->}
\Omega^{b\, A\dot{A}} = 4i e^{\alpha A} \chi_a{}^{\check{A}}_\alpha  + e^b
K_b{}^{a\; A\check{A}}
\; , \qquad {} \qquad \hat{E}^{\check{A}}_\alpha¨:=\hat{E}^{\underline{\alpha}}v_{\underline{\alpha}}{}^{\check{A}}_\alpha
= e^a \chi_a{}^{\check{A}}_\alpha \;  .
\end{eqnarray}
Notice that, by definition, $K_{ab}^{A\check{A}}= -
{\cal D}_{a}E_{b}^{\underline{a}}\; u_{\underline{a}}^{A\check{A}}$ so that one of the results of Eq. (\ref{DSemb=})
reads ${\cal D}_{[a}E_{b]}^{\underline{a}}\; u_{\underline{a}}^{A\check{A}}=0$. To see that this is natural, one can recall that in the linear approximation and in  the pure bosonic limit  $E_{b}^{\underline{a}}\mapsto  \partial_{b}\hat{x}^{\underline{a}}$. This also shows that the dynamical equations for the embedding function of super-5-brane can be formulated as an expression for the trace of $K_{ab}^{A\check{A}}$ tensor ($K_{a}{}^{a\; A\check{A}}= \partial_a\partial^a \hat{x}^{A\check{A}}+... $).

Taking into account the superembedding equation (\ref{EAdA=0}), as well as  Eq. (\ref{Du=uOm}) and (\ref{ea=Eua}), we find that the bosonic torsion of the worldvolume superspace is
\begin{eqnarray}\label{Dea=}
{\cal D}e^a= T^{\underline{a}} u_{\underline{a}}^a= -i e^{\alpha A} \wedge e^{\beta B}\epsilon_{AB}\gamma^a_{\alpha\beta} + i e^c\wedge e^b \epsilon_{\check{A}\check{B}}\chi_b^{\check{A}}\tilde{\gamma}^{a} \chi_c^{\check{B}} \;  .
\end{eqnarray}
This shall be used in studying the integrability conditions for the second equation in (\ref{DSemb->}), ${\cal D}(\hat{E}^{\check{A}}_\alpha¨-e^a \chi_a{}^{\check{A}}_\alpha)=0$. The dimension 1/2 com{ponent of this differential form equation (which appears as a coefficient for dim 1 two form $e^{\beta B}\wedge e^{\gamma C}$) gives the fermionic equation of motion of the simple $D=10$, ${\cal N}=1$ five-brane,
\begin{eqnarray}\label{fEqm=}
\tilde{\gamma}^{a \alpha\beta} \chi_{a}{}_\beta^{\check A} =0\;  .
\end{eqnarray}
One can check that its linearized limit is the massless d=6 Dirac equation for the Goldstone fermion of the five-brane,
$\tilde{\gamma}^{a \alpha\beta} \partial_{a}\hat{\theta}{}_\beta^{\check A} =0$.

Then, using (\ref{Tf=10D}) one finds that the dim 1 ($\propto e^{b}\wedge e^{\beta B}$) component of ${\cal D}(\hat{E}^{\check{A}}_\alpha¨-e^a \chi_a{}^{\check{A}}_\alpha)=0$ results in
$K_{ba\; B\check A}\gamma^a_{\alpha\beta}=-{i\over 2} h_{\underline{a}_1\underline{a}_2\underline{a}_3} \, v_{\beta B}\sigma^{\underline{a}_1\underline{a}_2\underline{a}_3}v_{\check A}{}^{\gamma }\gamma_{b\gamma \alpha}+ 2 D_{\beta B}\chi_{a\alpha \check A}$. Contracting this with $\tilde{\gamma}^{b\alpha\beta}$, and using the fermionic equation of motion (\ref{fEqm=}) as well as
the identity (\ref{vs3v'=}), one finds the bosonic equation of motion of the simple $D=10$, ${\cal N}=1$ five-brane,
\begin{eqnarray}\label{bEqm=}
\eta^{bc}K_{bc\; B\check A}:= - D^c \hat{E}_c{}^{\underline{a}}u_{\underline{a}B\check A}=
{3i\over 2} h_{\underline{a}\underline{b}\underline{c}}(\hat{Z})\, u_{B \check C}^{\underline{a}} u^{\underline{b} C \check C}u_{C \check A}^{\underline{c}}
\;  ,
\end{eqnarray}
as well as the following restriction on the pull--back $ \hat{h}_{\underline{a}\underline{b}\underline{c}} := h_{\underline{a}\underline{b}\underline{c}} (\hat{Z})$ of the Kalb-Ramond flux $h_{\underline{a}\underline{b}\underline{c}} ({Z})$  to the worldvolume superspace ${\cal W}{}^{(6|8)}$,
\begin{eqnarray}\label{huuu=}
 h_{\underline{a}\underline{b}\underline{c}} (\hat{Z}) u_{a}^{\underline{a}} u_b^{\underline{b} }u_{A \check A}^{\underline{c}} =0
\;  .
\end{eqnarray}

Hence,  the superembedding equation for a $D=10$, ${\cal N}=1$ super-5-brane contains the above described equations of motion among its consequences. It also determines completely (modulo freedom in choosing conventional constraints) the geometry of the worldline superspace which is thus induced by the (super)embedding. (See Appendix for some details).

\section{Towards the superembedding description of $SO(32)$ heterotic five-brane }

The story of heterotic 5-branes begins from the study of supersymmetric supergravity solutions
\cite{Duff94}. In addition to string solution solution representing a heterotic string, the
$D=10$ ${\cal N}=1$ supergravity has a 5-brane solution which can be considered as dual to the
string in the same way as magnetic monopole solution is dual to a solution describing a charged
particle in Electrodynamics (see \cite{Duff94} for review and original references).
The natural idea was that this corresponds to the $D=10$ ${\cal N}=1$ super-5-brane of the
'brane scan' of \cite{AETW} the worldvolume action of which is given by the sum of the the Nambu-Goto
and the Wess--Zumino terms  \cite{AETW}.
However, this action suffers from anomalies and is not the (complete) action of the heterotic 5-brane; to describe this additional worldvolume fields have to be introduced.

It was argued in \cite{Witten:1995gx} that the spectrum of SO(32) heterotic five-brane  contains, in addition to 'geometrical sector' $\hat{Z}^{\underline{\cal M}}(\xi) =( \hat{x}{}^{
\underline{m}}(\xi)\,  , \hat{\theta}^{ \underline{\mu}}(\xi))$ describing the embedding of 6-dimensional worldvolume $W^6$ into the target $D=10$ ${\cal N}=1$ superspace $\Sigma^{(10|16)}$, the 6-dimensional $N=2$ $SU(2)$ SYM multiplet and the hypermultiplet  in {\bf (2,32)} representation of $SU(2)\times SO(32)$. These are described by, respectively, a traceless $2\times 2$ matrix one form
 $A_{\tilde{B}}{}^{\tilde{A}}=d\xi^m A_{m \tilde{B}}{}^{\tilde{A}}$ ($A_{\tilde{B}}{}^{\tilde{B}}=0$, ${\tilde{A}}, {\tilde{B}}=1,2$) and by the set of bosonic scalar and fermionic spinor fields
 $(H^{A\tilde{B}J}(\xi), \psi_\alpha^{\tilde{B}J}(\xi))$
  related by supersymmetry transformations
  \begin{eqnarray}
\label{susyH=psi}
  \delta_{susy}H^{A\tilde{B}J}&=& 4i\epsilon^{\alpha A}\psi_\alpha^{\tilde{B}J}\; , \qquad \\ \nonumber  && A,B=1,2\; , \quad \alpha=1,2,3,4\; , \qquad {\tilde{A}}, {\tilde{B}}=1,2\; , \qquad J=1,\ldots , 32\; .
 \end{eqnarray}

The presence of additional worldvolume supermultiplets  makes the heterotic 5-brane similar to a  multi-p-brane system, the superembedding approach to which was proposed and developed for the case of multiple D$0$-branes (mD$0$) in \cite{IB09:D0} and for the case of and multiple M$0$-brane (mM$0$) in \cite{mM0=PLB,mM0=PRD}.

This experience of the previous study of mD$0$ and mM$0$ systems suggests the following proposition for the description of SO(32) heterotic 5-brane in the frame of superembedding approach. Let us introduce the $SU(2)$ connection one form on the $d=6$, $ n=2$ dimensional worldvolume superspace ${\cal W}^{(6|8)}$
\begin{eqnarray}
\label{SU(2)A=}
A_{\tilde{B}}{}^{\tilde{A}}= e^{\alpha C} A_{\alpha C\; \tilde{B}}{}^{\tilde{A}}(\zeta)+ e^a A_{a\; \tilde{B}}{}^{\tilde{A}}(\zeta)\; , \qquad  (A_{\tilde{B}}{}^{\tilde{A}})^* =- A_{\tilde{A}}{}^{\tilde{B}}
\quad (\Rightarrow \; A_{\tilde{A}}{}^{\tilde{A}}=0)\; ,  \end{eqnarray}
and the superfield $H^{A\tilde{B}J}(\zeta)$ in {\bf (2,32)} representation of $SU(2)\times SO(32)$. We restrict the field strength of the above connection by the constraints
\begin{eqnarray}
\label{SU(2)F=}
F_{\tilde{B}}{}^{\tilde{A}}:= (dA- A\wedge A)_{\tilde{B}}{}^{\tilde{A}}= {i\over 2}
e^a\wedge e^{\alpha {A}} \gamma_{b \alpha\beta} (W^{\beta}_A)_{\tilde{B}}{}^{\tilde{A}} +  {1\over 2}
e^b\wedge e^a (F_{ab})_{\tilde{B}}{}^{\tilde{A}}\; , \qquad  \end{eqnarray}
and the superfield $H^{A\tilde{B}J}(\zeta)$ by the equation
 \begin{eqnarray}
\label{DfH=psi}
  {\cal D}_{\gamma C} H^{A\tilde{B}J}= 4i\delta_C{}^{A}\psi_\gamma^{\tilde{B}J}\;  \qquad
 \end{eqnarray}
in which ${\cal D}_{\gamma C}$ is the fermionic $SO(1,5)\otimes SO(4)\otimes SU(2)=
SU(4)^*\otimes SU(2)\otimes SU(2)\otimes SU(2)$ covariant derivative on the worldvolume superspace ${\cal W}^{(6|8)}$ the geometry of which is assumed to be induced by its embedding in the $D=10$ ${\cal N}=1$ superspace $\Sigma^{(10|16)}$.

It is natural to propose (at least as an approximation) that the embedding  ${\cal W}^{(6|8)} \subset \Sigma^{(10|16)}$ is governed by the superembedding equation (\ref{SembEq=}),
\begin{eqnarray}
\label{SembEq=h}
  \hat{E}_{\beta B}{}^{\underline{a}}:=  {\cal D}_{\beta B}\hat{Z}^{\underline{\cal M}}\;  E_{\underline{\cal M}}{}^{\underline{a}}(\hat{Z}) = 0\; .
\qquad
\end{eqnarray}
Then (in this approximation and with our conventional constraints) the dynamics of the geometric sector is determined by equations of motion  (\ref{fEqm=}) and (\ref{bEqm=}),  which follow from the superembedding equation.

Hence, in such a (probably approximate) superembedding description  the geometric sector of the SO(32) heterotic five-brane, describing the embedding of the worldvolume ${W}^{6}$  in the target superspace $ \Sigma^{(10|16)}$, would obey the equations which can be obtained by varying the simple Nambu--Goto plus Wess--Zumino action of \cite{AETW}, while the characteristic properties of the $SO(32)$ heterotic 5-brane are carried by additional hypermultiplets and SU(2) SYM multiplet living on its worldvolume.  Our proposition is to describe their dynamics by ${\cal W}^{(6|8)}$  superfields obeying Eqs. (\ref{SU(2)F=}) and (\ref{DfH=psi}).
The consistency of these equations  on the worldvolume superspace  ${\cal W}^{(6|8)}$  is presently under investigation. Notice that, in our present settings, this superspace can be identified with the worldvolume superspace of  `simple' five-brane so that its geometry is described by equations presented in Sec. 2 and in Appendix.

Certainly, one may think about modification of the basic superembedding equations and also about considering the five-brane in the supergravity  and $SO(32)$ SYM background described by more general superspace constraints \cite{10D1N=SG+}. However, the natural first stage is to check first the consistency of our proposition, givin essentially by Eqs. (\ref{SU(2)F=}), (\ref{DfH=psi}) and (\ref{SembEq=h}), which, if consistent, would provide (at least) an approximate description of such a hypothetical more complete formulation of the SO(32) heterotic five-brane.

\begin{acknowledgements}
The author is thankful to Dima Sorokin, Mario Tonin and Paolo Pasti
for the interest to this work and for their warm hospitality at the
Padova Section of INFN and Padova University on its initial stages.
The partial support  by the research grants FIS2008-1980 from the
Spanish MICINN  and  by the Basque Government Research Group Grant ITT559-10 is greatly acknowledged.
\end{acknowledgements}

\appendix{{\bf Appendix: Geometry of the worldvolume superspace induced by the (super)embedding}}

As a consequence of the superembedding equation (after taking into account our conventional constraints and the target superspace supergravity constraints)  the bosonic torsion of the worldline superspace is determined by Eq. (\ref{Dea=}), while the fermionic torsion and the curvatures of ${\cal W}^{(6|8)}$ and of the normal bundle over it are given by
\begin{eqnarray}
\label{Def=}
{\cal D}e^{\alpha A}&=& e^b \wedge e^{\beta B} t_{\beta B\; b}{}^{\alpha A}  + {1\over 2} e^b \wedge e^{a} t_{a b}{}^{\alpha A}
\; ,  \qquad \nonumber \\  t_{\beta B\; b}{}^{\alpha A}&=& 2i \chi_{a\beta \check B}\chi_{b\gamma}{}^{\check B}\tilde{\gamma}{}^{a\gamma\alpha}-{i\over 4} \hat{h}_{c_1c_2c_3} (\gamma^{c_1c_2c_3}\gamma_b)_\beta{}^\alpha \delta_B{}^A -{3i\over 4} \hat{h}_{b\, B\check{B}}{}^{A\check{B}} (\gamma^{a}\gamma_b)_\beta{}^\alpha \delta_B{}^A \; , \nonumber \\ t_{a b}{}^{\alpha A}&=& \epsilon_{\check{B}\check{C}}(\chi_{[a|}{}^{\check C}\tilde{\gamma}{}^c){}^\alpha K_{|b]c}{}^{A\check B}+ {i\over 2} \hat{h}_{D\check C}{}^{D\check A\; A\check C} \epsilon_{\check{A}\check{B}} (\chi_{[a}^{\check B}\tilde{\gamma}_{b]})^\alpha  + \qquad \nonumber \\ && + {3i\over 2} \epsilon^{{A}{B}} \hat{h}_{cd\; B\check B} (\tilde{\gamma}_{[a}\gamma^{cd}\chi_{b]}^{\check B})^\alpha  +\hat{T}{}_{a b}{}^{\alpha A} \qquad \; ,
\\
\label{rab=}
r^{ab} &=& \hat{R}{}^{ab} + 8 e^{\alpha A}\wedge e^{\beta B} \epsilon_{AB}\epsilon_{\check{A}\check{B}} \chi_{\alpha}^{a\check A} \chi_{\beta}^{b\check B} - 4i e^{c}\wedge e^{\alpha A} \chi_{\alpha}^{[a|\check B}K_{c}{}^{|b]}{}_{A \check A} + \qquad \nonumber \\ && + {1\over 2} e^c \wedge e^{d} K_{c}{}^{a}{}_{A \check A}K_{d}{}^{b\; A \check A}\; , \quad \\
 \label{FBA=}
{\cal F}_{B}{}^{A} &=& \hat{R}{}^{ab} + 8 e^{\alpha A}\wedge e^{\beta B} \epsilon_{AB}\epsilon_{\check{A}\check{B}} \chi_{\alpha}^{a\check A} \chi_{\beta}^{b\check B} - 4i e^{c}\wedge e^{\alpha A} \chi_{\alpha}^{[a|\check B}K_{c}{}^{|b]}{}_{A \check A} + \qquad \nonumber \\ && + {1\over 2} e^c \wedge e^{d} K_{c}{}^{a}{}_{A \check A}K_{d}{}^{b\; A \check A}\; . \qquad
\end{eqnarray}

\end{document}